%% file: main.tex
\documentclass[smallabstract,smallcaptions]{dccpaper}

\usepackage{citesort}
\usepackage{amsmath}
\usepackage{amssymb}
\usepackage{color}
\usepackage{url}
\usepackage[capitalise]{cleveref}
\usepackage{booktabs}
\usepackage{tikz}
\usetikzlibrary{positioning}

\usepackage[utf8]{inputenc}
\usepackage{pgfplots}

\newlength{\figurewidth}
\newlength{\smallfigurewidth}

\tikzset{font={\fontsize{10pt}{12}\selectfont}}

\title{\large
  \textbf{Predicting Chroma from Luma in AV1}
}

\author{%
  Luc N. Trudeau$^{\ast\dag}$, Nathan E. Egge$^{\ast\dag}$, and David Barr$^{\dag}$\\[0.5em]
  {\small\begin{minipage}{\linewidth}\begin{center}
    \begin{tabular}{ccc}
    $^{\ast}$Mozilla & \hspace*{0.5in} & $^{\dag}$Xiph.Org Foundation \\
      331 E Evelyn Ave  && 21 College Hill Road \\
      Mountain View, CA, 94041, USA && Somerville, MA, 02144, USA\\
      \url{luc@trud.ca}, \url{negge@mozilla.com} && \url{b@rr-dav.id.au}
    \end{tabular}
    \end{center}\end{minipage}}
}

\newcommand{\mpred}[1]{#1^{\text{p}}}
\newcommand{\mrecon}[1]{#1^{\text{r}}}
\newcommand{\mz}[1]{#1^{\text{AC}}}
\newcommand{\msumsum}{\sum_{i}\sum_{j}}
\newcommand{\msubsum}{\text{S}}

\begin{document}

\maketitle

\begin{abstract}
Chroma from luma~(CfL) prediction is a new and promising chroma-only intra
predictor that models chroma pixels as a linear function of the coincident
reconstructed luma pixels. In this paper, we present the CfL predictor adopted
in Alliance Video 1 (AV1), a royalty-free video codec developed by the Alliance
for Open Media (AOM). The proposed CfL distinguishes itself from prior art not
only by reducing decoder complexity, but also by producing more accurate
predictions. On average, CfL reduces the BD-rate, when measured with CIEDE2000,
by 4.87\% for still images and 2.41\% for video sequences.
\end{abstract}

\Section{Introduction}

Still image and video compression is typically not performed using red, green,
and blue~(RGB) color primaries, but rather with a color space that separates
luma from chroma. There are many reasons for this, notably that luma and chroma
are less correlated than RGB, which favors compression; and also that the human
visual system is less sensitive to chroma allowing one to reduce the resolution
in the chromatic planes, a technique known as chroma subsampling~\cite{Wang01}.

Another way to improve compression in still images and videos is to subtract
the pixels by a predictor. When this predictor is derived from previously
reconstructed information inside the current frame, it is referred to as an
intra prediction tool. In contrast, an inter prediction tool uses information
from previously reconstructed frames. For example, ``DC'' prediction is an
intra prediction tool that predicts the pixels values in a block by averaging
the values of neighboring pixels adjacent to the above and left borders of the
block~\cite{Li14}.

Chroma from luma~(CfL) prediction is a new and promising chroma-only intra
predictor that models chroma pixels as a linear function of the coincident
reconstructed luma pixels~\cite{Kim10}. It was proposed for the HEVC video
coding standard~\cite{Chen11b}, but was ultimately rejected, as the decoder
model fitting caused a considerable complexity increase. It was proposed again
as part of the HEVC Range Extension~\cite{Pu13}, this time without decoder
model fitting in order to reduce decoder complexity.

More recently, CfL prediction was implemented in the Thor
codec~\cite{Midtskogen16} as well as in the Daala codec~\cite{Egge15}. The
inherent conceptual differences in the Daala codec, when compared to HEVC, led
to multiple innovative contributions by Egge and Valin~\cite{Egge15} to CfL
prediction. Most notably a frequency domain implementation.

As both Thor and Daala served as bases for AV1, a research initiative was
established regarding CfL, the results of which are presented in this paper.
The proposed CfL implementation, outlined in \cref{sec:outline}, not only
builds on the innovations of~\cite{Egge15}, but does so in a way that is
compatible with the more conventional compression tools found in AV1. This new
implementation is considerably different from its predecessors. Its key
contributions are:
\begin{itemize}
  \item Enhanced parameter signaling, as described in \cref{sec:Signaling},
when compared with~\cite{Pu13}, the proposed signaling is more precise and
finding the RD-optimal parameters is less complex.
  \item Model fitting the ``AC'' contribution of the reconstructed luma pixels,
as shown in~\cref{sec:ZeroMean}, which simplifies the model and allows for a
more precise fit.
  \item Chroma ``DC'' prediction for ``DC'' contribution, which requires no
signaling and, as described in~\cref{sec:DC_PRED}.
\end{itemize}
Finally,~\cref{sec:Results} presents detailed results of the compression gains
of the proposed CfL prediction implementation in AV1.

\Section{State of the Art in Chroma from Luma Prediction}
\label{sec:StateOfTheArt}

As described in~\cite{Kim10}, CfL prediction models chroma pixels as a linear
function of the coincident reconstructed luma pixels. More precisely, let $L$
be an $M\times N$ matrix of pixels in the luma plane; we define $C$ to be the
chroma pixels spatially coincident to $L$. Since $L$ is not available to the
decoder, the reconstructed luma pixels, $\mrecon{L}$, corresponding to $L$ are
used instead. The chroma pixel prediction, $\mpred{C}$, produced by CfL uses
the following linear equation:
\begin{equation}
  \mpred{C} = \alpha \times \mrecon{L} + \beta \:.
\end{equation}

Some implementations of CfL~\cite{Kim10,Chen11b,Midtskogen16} determine the
linear model parameters $\alpha$ and $\beta$ using linear least-squares
regression
\begin{equation}
  \alpha = \frac
    {(M\times N) \msumsum \mrecon{L}_{ij} C_{ij} - \msumsum \mrecon{L}_{ij} \msumsum C_{ij}}
    {(M\times N) \msumsum {(\mrecon{L}_{ij})}^2 - {(\msumsum \mrecon{L}_{ij})}^2}\:,
\end{equation}
\begin{equation}
  \beta = \frac
  {\msumsum C_{ij} - \alpha \msumsum \mrecon{L}_{ij}}
  {M \times N} \:.
\end{equation}

We classify~\cite{Kim10,Chen11b,Midtskogen16} as implicit implementations of
CfL, since $\alpha$ and $\beta$ are not signaled in the bitstream, but are
implied from the bitstream. The main advantage of the implicit implementation
is the absence of signaling.

However, implicit implementations have numerous disadvantages. As mentioned
before, computing least squares considerably increases decoder complexity.
Another important disadvantage is that the chroma pixels, $C$, are not
available when computing least squares on the decoder.  As such, prediction
error increases since neighboring reconstructed chroma pixels must be used
instead.

In~\cite{Egge15}, the authors argue that the advantages of explicit signaling
considerably outweigh the signaling cost. This is corroborated by the results
in~\cite{Pu13}. Based on these findings, we propose a hybrid approach that
signals $\alpha$ and implies $\beta$.

\Section{The Proposed Chroma from Luma Prediction}
\label{sec:outline}

This section outlines the proposed chroma-only intra predictor. To predict
chroma samples, CfL starts with the spatially coinciding reconstructed luma
pixels.

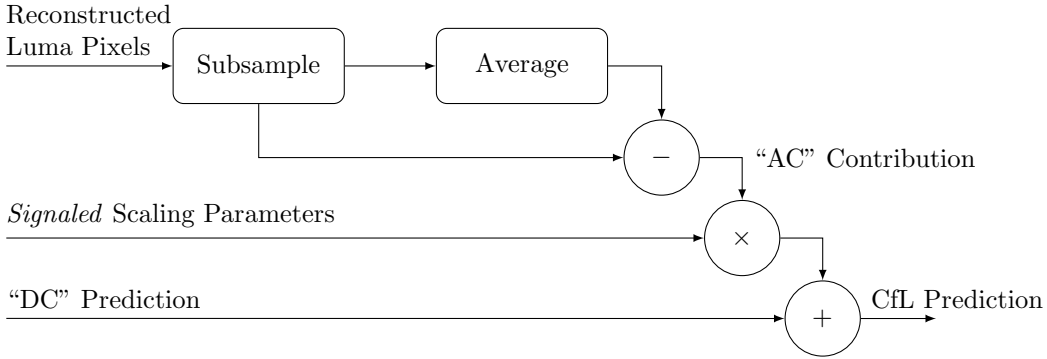
\begin{figure}[htb]
  \centering
  \begin{tikzpicture}[node distance = 3.5cm, auto]
    \tikzstyle{block} = [rectangle, draw, text width=2cm, text centered,
    rounded corners, minimum height=1cm]
    \tikzstyle{op} = [circle, draw, text width=0.45cm, text centered, minimum height=1cm]
    \node [] (start) {};
    \node [below = 2cm of start] (scaling) {};
    \node [below = 0.78cm of scaling] (dc) {};
    \node [block, right of = start] (sub) {Subsample};
    \node [block, right of = sub] (avg) {Average};
    \node [op, below right = 0.5cm of avg] (minus) {$-$};
    \node [op, below right = 0.5cm of minus] (times) {$\times$};
    \node [op, below right = 0.5cm of times] (plus) {$+$};

    \node [right = 1cm of plus] (end) {};
    \draw[-latex] (start) node[above right, text width=4cm] {Reconstructed\newline Luma Pixels} -> (sub);
    \draw[-latex] (sub) -> (avg);
    \draw[-latex] (avg) -| (minus);
    \draw[-latex] (sub) |- (minus);
    \draw[-latex] (minus) -| node{``AC'' Contribution} (times);
    \draw[-latex] (times) -| (plus);
    \draw[-latex] (scaling) node[above right, text width=6cm] {\textit{Signaled} Scaling Parameters} -> (times);
    \draw[-latex] (dc) node[above right, text width=6cm] {``DC'' Prediction} -> (plus);
    \draw[-latex] (plus) node[above right = 0cm and 0.5cm, text width=3cm] {CfL Prediction} -> (end);

  \end{tikzpicture}
  \caption{Outline of the operations required to build the proposed CfL prediction.}\label{fig:overview}
\end{figure}

As illustrated in \cref{fig:overview}, when chroma subsampling is used, in
order for the pixels to coincide, the reconstructed luma pixels are subsampled
accordingly. As explained in~\cref{sec:ZeroMean}, the coinciding reconstructed
luma pixels are subtracted by their average, which results in their ``AC''
contribution.

As for the scaling factor indices and signs, they are decoded from the
bitstream, which is described in~\cref{sec:Signaling}. The CfL prediction is
built by multiplying the ``AC'' contribution of reconstructed luma pixels with
the scaling factors and the result is added to the intra ``DC'' prediction, as
explained in~\cref{sec:DC_PRED}.

\Section{Model Fitting the ``AC'' Contribution}
\label{sec:ZeroMean}

In~\cite{Egge15}, Egge and Valin demonstrate the merits of separating the
``DC'' and ``AC'' contributions of the frequency domain CfL prediction. In the
pixel domain, the ``AC'' contribution of a block can be obtained by subtracting
it by its average.

An important advantage of the ``AC'' contribution is that it is zero mean,
which results in significant simplifications to the least squares model
parameter equations. More precisely, let $\mz{L}$ be the zero-meaned
reconstructed luma pixels. Because $\msumsum \mz{L} = 0$, substituting
$\mrecon{L}$ by $\mz{L}$ yields the following simplified model parameters
equations:
\begin{equation}
  \mz{\alpha} = \frac{\msumsum \mz{L}_{ij} C_{ij}}{\msumsum {(\mz{L}_{ij})}^2} \:,
\end{equation}
\begin{equation}
  \mz{\beta} = \frac{\msumsum C_{ij}}{M \times N} \:.
  \label{eq:beta}
\end{equation}
We define the zero-mean chroma prediction, $\mz{C}$, like so
\begin{equation}
  \mz{C} = \mz{\alpha} \times \mz{L} + \mz{\beta} \:.
\end{equation}

When computing the zero-mean reconstructed pixels, the resulting values are
stored using $1/8$th precision fixed-point values. This ensures that even with
12-bit integer pixels, the average can be stored in a 16-bit signed integer.

Traditionally, subsampling is performed by adding the coincident pixels in the
luma plane and dividing by the number of pixels. The exact number of coincident
pixels is determined by the type subsampling. AV1 supports: 4:2:0, 4:2:2, 4:4:0
and 4:4:4 chroma subsamplings~\cite{Wang01}. Let $s_x, s_y \in \{1,2\}$ be the
subsampling steps along the x and y axes, respectively. It follows that the
summing the coincident pixels at position $(u,v)$ is performed as follows:
\begin{equation}
  \msubsum(s_x, s_y, u, v) = \sum_{y=1}^{s_y} \sum_{x=1}^{s_x}
        \mrecon{L}_{s_{y}\times v + y, s_{x}\times u + x}
\end{equation}

This luma subsampling step was considered too costly for HEVC~\cite{Chen11b}
which explains why~\cite{Pu13} is only available for 4:4:4. We propose a simpler
subsampling scheme that is less complex and more precise.

By combining the luma subsampling step with the average subtraction step (shown
in \cref{fig:overview}), not only do the equations simplify, but the
subsampling divisions and the corresponding rounding error are removed. The
equation corresponding to the combination of both steps is given in
\cref{eq:Q3}, which simplifies to \cref{eq:Q3Simple}. Note that both equations
use integer divisions.
\begin{equation}
  \mz{L}_{u,v} = 8 \left(
     \frac{\msubsum(s_x, s_y, u, v)} {s_y \times s_x}
  \right) - \frac{8\msumsum \left(
     \frac{\msubsum(s_x, s_y, i, j)}{s_y \times s_x}
 \right)}
{M \times N}
\label{eq:Q3}
\end{equation}
\begin{equation}
\implies \frac{1}{s_y \times s_x}\left(
    8 \times \msubsum(s_x, s_y, u, v)
    - \frac{\msumsum 8 \times \msubsum(s_x, s_y, i, j) }{M \times N}
    \right)
\label{eq:Q3Simple}
\end{equation}

Based on the supported chroma subsamplings, it can be shown that $s_y \times
s_x \in \{ 1,2,4\}$ and that since both $M$ and $N$ are powers of two, $M
\times N$ is also a power of two. It follows that both $\frac{1}{s_y \times
s_x}$ and $\frac{1}{M \times N}$ in \cref{eq:Q3Simple} can be replaced by bit
shift operations.

For example, in the context of a 4:2:0 chroma subsampling, instead of applying
a box filter, the proposed approach only requires to sum the 4 reconstructed
luma pixels that coincide with the chroma pixels. Afterwards, when CfL will
scale its luma pixels to improve the precision of the predictions,~\cite{Pu13}
requires to scale by 8, whereas the proposed approach only needs to scale by 2
(i.e. $\frac{8}{2 \times 2}$).  Both approach are now on the same scale but the
rounding errors saved in~\cref{eq:Q3Simple} results in more precise values for
the proposed approach. In other words, we will perform the integer division
required by the box filter only when we downscale the predicted pixel values.

\Section{Chroma ``DC'' Prediction for ``DC'' Contribution}
\label{sec:DC_PRED}

Switching the linear model to use zero mean reconstructed luma pixels also
changes the ``DC'' contribution, to the extent that it now only depends on $C$.
This can be seen in \cref{eq:beta}, where $\mz{\beta}$ is the average of the
chroma pixels.

\begin{figure}[htb]
\centering
\input{dcpred_err_boxplot.tex}
\caption{Error analysis of the ``DC'' predictor.}\label{fig:dc_pred}
\end{figure}
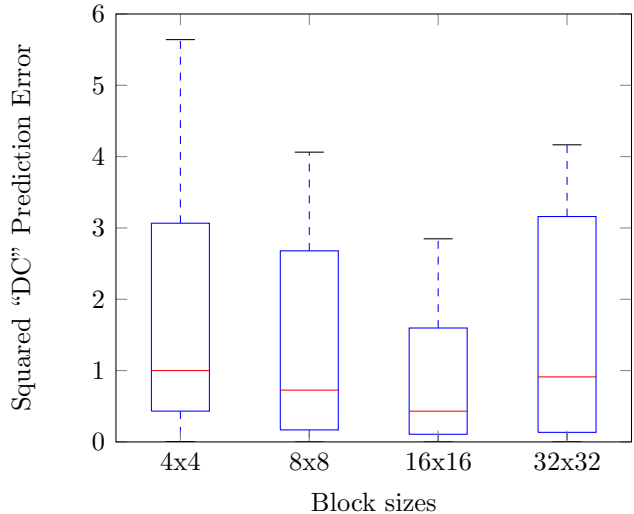

The chroma pixel average for a given block is not available in the decoder.
However, there already exists an intra prediction tool that predicts this
average. When applied to the chroma plane, the ``DC'' prediction predicts the
pixel values in a block by averaging the values of neighboring pixels adjacent
to the above and left borders of the block~\cite{Li14}. In~\cref{fig:dc_pred},
we present an analysis of the ``DC'' prediction error over the Kodak True Color
Image suite.

Note that \cref{fig:dc_pred} does not include outliers, as they hinder
readability of the figure. Furthermore, it is unlikely that the intra mode
selection algorithm would select CfL to predict such content. The median
squared ``DC'' prediction error is equal to or less than 1 for all tested block
sizes.

Based on this error analysis, the absence of signaling and the fact that ``DC''
prediction is already implemented in AV1, we use ``DC'' prediction as an
approximation for $\mz{\beta}$, as shown in~\cref{fig:overview}. The proposed
CfL prediction is expressed as follows:
\begin{equation}
\text{CfL}(\alpha) = \alpha \times \mz{L} + \text{DC} \:.
\end{equation}

\Section{Parameter Signaling}
\label{sec:Signaling}

Signaling the scaling parameters allows encoder-only fitting of the linear
model. This reduces decoder complexity and results in a more precise
prediction, as the best scaling parameter can be determined based on the
reference chroma pixels which are only available to the encoder. 

Signaling the scaling parameters fundamentally changes their selection. In this
context, the least-squares regression used in~\cite{Kim10,Chen11b,Midtskogen16}
does not yield an RD-optimal solution as it ignores the trade-off between the
rate and the distortion of the scaling parameters. For the proposed CfL
prediction, the signaling parameters are determined using the same
rate-distortion optimization mechanics as other coding tools and parameters of
AV1. Concretely, given a set of scaling parameters $A$, the selected scaling
parameter is the one that minimizes the trade-off between the rate and the
distortion

\begin{equation}
\alpha = \arg\min_{a \in A} \left( D(\text{CfL}(a)) + \lambda R(a) \right) \:.
\end{equation}

In the previous equation, the distortion, $D$, is the sum of the squared error
between the reconstructed chroma pixels and the reference chroma pixels.
Whereas, the rate, $R$, is the number of bits required to encode the scaling
parameter and the residual coefficients. Furthermore, $\lambda$ is the weighing
coefficient between rate and distortion used by AV1.

CfL parameters are signaled at the prediction unit level, with consideration to
the fact that rate-distortion optimization approaches are used over the
traditional least squares regression. Since CfL is an chroma-only intra
prediction mode, there is no need to always signal a skip flag, when CfL is not
desired as is the case for~\cite{Pu13}.

CfL parameters apply to the whole prediction unit. The ``DC'' prediction, used
by CfL, is computed over the entire prediction unit. This greatly reduces the
rate-distortion search space of CfL parameters as they do not interact with the
interdependencies of the transform blocks inside a prediction unit.

In~\cite{Pu13}, CfL parameter signaling is at the transform block level. This
creates more interdependencies between transform blocks as the CfL parameters
will change the pixels used when performing intra prediction on neighboring
transform blocks. Evaluating all these combinations is prohibitively expensive
resulting in the use of fast approximation approaches that cannot guarantee
optimal results. The proposed solution avoid these issues, with the added
benefit that reducing interdependencies between transform blocks speeds up the
prediction and reconstruction process of prediction units.

When the CfL chroma only mode is chosen, we first signal the joint sign of both
scaling parameters. A sign is either negative, zero, or positive. Contrary
to~\cite{Pu13}, the proposed signaling does not permit choosing (zero, zero),
as it results in ``DC'' prediction. It follows that the joint sign requires an
eight-value symbol.

As for each scaling parameter, a 16-value symbol is used to represent values
ranging from 0 to 2 with a step of $1/8$th. The entropy coding details are
beyond the scope of this paper; however, it is important to note that a
16-value symbol fully utilizes the capabilities of the multi-symbol entropy
encoder~\cite{Valin16}. In comparison with~\cite{Pu13}, the proposed signaling
scheme offers twice the range and twice the precision. Finally, scaling
parameters are signaled only if they are non-zero.

\Section{Experimental Results}
\label{sec:Results}

To ensure a valid evaluation of coding efficiency gains, our testing
methodology conforms to that of~\cite{Daede17}. All simulation parameters and a
detailed sequence-by-sequence breakdown for all the results presented in this
paper are available online at~\cite{AWCY}. Furthermore, the bitstreams
generated in these simulations can be retrieved and analyzed online
at~\cite{Analyzer}.

The following tables show the average percent rate difference measured using
the Bj{\o}ntegaard rate difference, also known as BD-rate~\cite{Bjontegaard01}.
The BD-rate is measured using the following objective metrics: PSNR,
PSNR-HVS~\cite{Egiazarian2006}, SSIM~\cite{Wang04}, CIEDE2000~\cite{Yang12} and
MSSIM~\cite{Wang03}. Of all the previous metrics, only the CIEDE2000 considers
both luma and chroma planes. It is also important to note that the distance
measured by this metric is perceptually uniform~\cite{Yang12}.

As required in~\cite{Daede17}, for individual feature changes in libaom, we use
quantizers: 20, 32, 43, and 55. We present results for three test sets:
Objective-1-fast~\cite{Daede17}, Subset1~\cite{testset} and
Twitch~\cite{testset}.

In \cref{tab:subset1}, we present the results for the Subset1 test set.  This
test set contains still images, which are ideal to evaluate the chroma intra
prediction gains of CfL when compared to other intra prediction tools in AV1.

\begin{table}
  \centering
  \caption{Results over the Subset1 test set (still images), available online~\cite{AWCYSubset1}.}\label{tab:subset1}
  \scriptsize
  \begin{tabular}{lrrrrrrr} \toprule
            & \multicolumn{7}{c}{BD-Rate} \\ \cmidrule(r){2-8}
            & PSNR & PSNR-HVS & SSIM & \textbf{CIEDE2000} & PSNR Cb & PSNR Cr & MS SSIM \\ \midrule
    Average & -0.53 & -0.31 & -0.34 & \textbf{-4.87} & -12.87 & -10.75 & -0.34 \\ \bottomrule
  \end{tabular}
\end{table}

For still images, when compared to all of the other intra prediction tools of
AV1 combined, CfL prediction reduces the rate by an average of 4.87\% for the
same level of visual quality measured by CIEDE2000.

For video sequences, \cref{tab:objective1} breaks down the results obtained
over the objective-1-fast test set.

\begin{table}
  \centering
  \caption{Results over the Objective-1-fast test set (video sequences), available online~\cite{AWCYObjective1}.}\label{tab:objective1}
  \scriptsize
  \begin{tabular}{lrrrrrrr} \toprule
            & \multicolumn{7}{c}{BD-Rate} \\ \cmidrule(r){2-8}
            & PSNR & PSNR-HVS & SSIM & \textbf{CIEDE2000} & PSNR Cb & PSNR Cr & MS SSIM \\ \midrule
    Average & -0.43 & -0.42 & -0.38 & \textbf{-2.41} & -5.85 & -5.51 & -0.40 \\ \midrule
    1080p   & -0.32 & -0.37 & -0.28 & \textbf{-2.52} & -6.80 & -5.31 & -0.31 \\ \midrule
    1080p-screen & -1.82 & -1.72 & -1.71 & \textbf{-8.22} & -17.76 & -12.00 & -1.75\\ \midrule
    360p    & -0.15 & -0.05 & -0.10 & \textbf{-0.80} & -2.17 & -6.45 & -0.04 \\ \midrule
    720p    & -0.12 & -0.11 & -0.07 & \textbf{-0.52} & -1.08 & -1.23 & -0.12 \\ \bottomrule
  \end{tabular}
\end{table}

Not only does CfL yield better intra frames, which produces a better reference
for inter prediction tools, but it also improves chroma intra prediction in
inter frames. We observed CfL predictions in inter frames when the predicted
content was not available in the reference frames. As such, CfL prediction
reduces the rate of video sequences by an average of 2.41\% for the same level
of visual quality when measured with CIEDE2000.

The average rate reductions for 1080p-screen are considerably higher than those
of other types of content. This indicates that CfL prediction considerably
outperforms other AV1 predictors for screen content coding. As shown in
table~\ref{tab:twitch}, the results on the Twitch test set, which contains only
gaming-based screen content, corroborates this finding.

\begin{table}
  \centering
  \caption{Results over the Twitch test set (gaming screen content), available online~\cite{AWCYTwitch}.}\label{tab:twitch}
  \scriptsize
  \begin{tabular}{lrrrrrrr} \toprule
            & \multicolumn{7}{c}{BD-Rate} \\ \cmidrule(r){2-8}
            & PSNR & PSNR-HVS & SSIM & \textbf{CIEDE2000} & PSNR Cb & PSNR Cr & MS SSIM \\ \midrule
    Average & -1.01 & -0.93 & -0.90 & \textbf{-5.74} & -15.58 & -9.96 & -0.81 \\ \bottomrule
  \end{tabular}
\end{table}

The sequence-by-sequence results presented in~\cite{AWCYTwitch} indicate that
CfL prediction is particularly efficient for sequences of the game Minecraft,
where the average rate reduction exceeds 20\% for the same level of visual
quality measured by CIEDE2000.

\Section{Conclusion}

In this paper, we presented the chroma from luma prediction tool adopted in
AV1.  This new implementation is considerably different from its predecessors.
Its key contributions are: parameter signaling, model fitting the ``AC''
contribution of the reconstructed luma pixels, and chroma ``DC'' prediction for
``DC'' contribution.  Not only do these contributions reduce decoder
complexity, but they also reduce prediction error; resulting in a 4.87\%
average reduction in BD-rate, when measured with CIEDE2000, for still images,
and 2.41\% for video sequences. Possible improvements to the proposed solution
includes non-linear prediction models and motion-compensated CfL.

\section*{Reference to Prior Literature}
%

\bibliographystyle{IEEEtran}
\bibliography{refs}
\end{document}

%% file: dcpred_err_boxplot.tex
\begin{tikzpicture}

\begin{axis}[
xlabel={Block sizes},
ylabel={Squared ``DC'' Prediction Error},
xmin=0.5, xmax=4.5,
ymin=0, ymax=6,
axis on top,
xtick={1,2,3,4},
xticklabels={4x4,8x8,16x16,32x32},
ytick={0,1,2,3,4,5,6},
tick pos=both
]
\addplot [blue]
table {%
0.775 0.431640625
1.225 0.431640625
1.225 3.06640625
0.775 3.06640625
0.775 0.431640625
};
\addplot [blue, dashed]
table {%
1 0.431640625
1 0.00390625
};
\addplot [blue, dashed]
table {%
1 3.06640625
1 5.640625
};
\addplot [black]
table {%
0.8875 0.00390625
1.1125 0.00390625
};
\addplot [black]
table {%
0.8875 5.640625
1.1125 5.640625
};
\addplot [red]
table {%
0.775 1
1.225 1
};
\addplot [blue]
table {%
1.775 0.16827392578125
2.225 0.16827392578125
2.225 2.67889404296875
1.775 2.67889404296875
1.775 0.16827392578125
};
\addplot [blue, dashed]
table {%
2 0.16827392578125
2 0.000244140625
};
\addplot [blue, dashed]
table {%
2 2.67889404296875
2 4.062744140625
};
\addplot [black]
table {%
1.8875 0.000244140625
2.1125 0.000244140625
};
\addplot [black]
table {%
1.8875 4.062744140625
2.1125 4.062744140625
};
\addplot [red]
table {%
1.775 0.7266845703125
2.225 0.7266845703125
};
\addplot [blue]
table {%
2.775 0.107257843017578
3.225 0.107257843017578
3.225 1.59715270996094
2.775 1.59715270996094
2.775 0.107257843017578
};
\addplot [blue, dashed]
table {%
3 0.107257843017578
3 0.0001373291015625
};
\addplot [blue, dashed]
table {%
3 1.59715270996094
3 2.84765625
};
\addplot [black]
table {%
2.8875 0.0001373291015625
3.1125 0.0001373291015625
};
\addplot [black]
table {%
2.8875 2.84765625
3.1125 2.84765625
};
\addplot [red]
table {%
2.775 0.430908203125
3.225 0.430908203125
};
\addplot [blue]
table {%
3.775 0.134082078933716
4.225 0.134082078933716
4.225 3.16010308265686
3.775 3.16010308265686
3.775 0.134082078933716
};
\addplot [blue, dashed]
table {%
4 0.134082078933716
4 0.000916481018066406
};
\addplot [blue, dashed]
table {%
4 3.16010308265686
4 4.16574478149414
};
\addplot [black]
table {%
3.8875 0.000916481018066406
4.1125 0.000916481018066406
};
\addplot [black]
table {%
3.8875 4.16574478149414
4.1125 4.16574478149414
};
\addplot [red]
table {%
3.775 0.911501407623291
4.225 0.911501407623291
};
\end{axis}

\end{tikzpicture}